\documentclass{article}

\usepackage{arxiv}

\usepackage[utf8]{inputenc} 
\usepackage[T1]{fontenc}    
\usepackage{hyperref}       
\usepackage{url}            
\usepackage{booktabs}       
\usepackage{amsfonts}       
\usepackage{nicefrac}       
\usepackage{microtype}      
\usepackage{lipsum}
\usepackage{graphicx}
\graphicspath{ {./images/} }
\usepackage{amsmath}
\usepackage{xcolor}

\title{Learning phase diversity for solving ill-posed inverse problems in imaging}

\author{
  Jasleen Birdi \\
  Optics and Photonics Centre \\
  Indian Institute of Technology Delhi \\
  New Delhi 110016 INDIA
   \And
 Tamal Majumder \\
  Department of Physics\\
  Indian Institute of Technology Delhi \\
  New Delhi 110016 INDIA\\
  \And
  Debanjan Halder \\
  Department of Physics\\
  Indian Institute of Technology Delhi \\
  New Delhi 110016 INDIA\\
  \And
  Muskan Kularia \\
  Optics and Photonics Centre \\
  Indian Institute of Technology Delhi \\
  New Delhi 110016 INDIA
  \And
  Kedar Khare \\
  Optics and Photonics Centre \\
  Indian Institute of Technology Delhi \\
  New Delhi 110016 INDIA \\
  kedark@opc.iitd.ac.in\\
}

\begin{document}
\maketitle
\begin{abstract}
Inverse problems in imaging are typically ill-posed and are usually solved by employing regularized optimization techniques.
The usage of appropriate constraints (e.g. image sparsity) can restrict the solution space, thus making it feasible for a reconstruction algorithm to find a meaningful solution even when the measured data are noisy and/or incomplete. In recent years, deep network based ideas aimed at learning the end-to-end mapping between the raw measurements and the target image have gained popularity. In the learning approach, it is not required to treat every instance of measured data as a fresh reconstruction problem. Instead, the functional relationship between the measured raw data and the solution image are learned by training a deep network with prior examples. While this approach allows one to significantly increase the real-time operational speed of a computational imaging system, it does not change the nature of the underlying ill-posed inverse problem. It is well-known in the inverse imaging community that availability of diverse non-redundant data via additional measurements can generically improve the robustness of the reconstruction algorithms. The multiple data measurements, however, typically demand additional hardware and complex system setups that are not desirable for field deployment. In this work, we note that in both incoherent and coherent optical imaging, the irradiance patterns corresponding to two phase diverse measurements associated with the same test object have implicit local correlation which may be learned. A physics informed data augmentation scheme is then described where a trained network is used for generating a phase diverse pseudo-data based on a ground truth data frame. The true data along with the augmented pesudo-data are observed to provide high quality inverse solutions with simpler reconstruction algorithms that are not sensitive to any parameter tuning like the traditional regularized optimization methods. We validate this approach for both incoherent and coherent optical imaging (or phase retrieval) configurations with vortex phase as a diversity mechanism. Our results may open new avenues for leaner high-fidelity computational imaging systems across a broad range of applications.
\end{abstract}


\section{Introduction}
Optical imaging systems, such as smartphone cameras and microscopes, are an integral part of modern life, capturing visual information from the world around us. These systems operate by transforming incident light into digital images through a sequence of physical and computational processes. Typically, the raw data collected by these devices, represented as $y$, is degraded by noise $n$ and affected by various systemic distortions or transformations (e.g., diffraction, aberrations, motion blur), which may be collectively modeled by an operator $A$ leading to a data generation model $y = A(x) + n$. Reconstructing a high-quality image $x$,  representing the original object from these measurements constitutes an inverse problem. Solution of the inverse problem is central to working of computational imaging systems, which leverage hardware improvements and algorithmic methods to recover the true scene from imperfect data, thereby enabling more accurate and informative visual representations in a wide range of applications. In the context of biophotonics applications, microscopy systems \cite{mertz} employing spatially incoherent and coherent illumination have become increasingly computational in nature. Incoherent microscopic imaging such as bright-field and fluorescence microscopy primarily benefits from computational methods such as denoising, deconvolution as well as super-resolution. Coherent or quantitative phase imaging in microscopy has also become important for label-free cell imaging applications including inline or off-axis digital holography and phase tomography.  

First highlighted by Hadamard \cite{Hadamard1923}, the reconstruction problems are inherently ill-posed due to noise, limited measurements, or inherent ambiguities. This ill-posedness makes image reconstruction both challenging and a continuing focus of research. {Over the past several decades, advances in imaging and computational hardware accompanied by new algorithms - from optimization to deep learning - have significantly improved the performance of computational imaging systems. Ill-posedness is however a fundamental issue that is independent of the algorithmic approach used for image reconstruction. It is generally accepted that having additional non-redundant measurements of the unknown object to be imaged can certainly improve the robustness of inverse algorithms. Such an approach involving multiple measurements often requires complex (or sometimes impossible) hardware setups that make such approaches undesirable for field deployment. In the case of incoherent and coherent optical imaging, the diverse measurements can for example be generated by means of a sequence of linear (e.g. diffraction, transmission through an optical element, etc.) and non-linear (e.g. detection)
processes. Our work is motivated from the observation that phase diverse measurements (as explained in more detail later in the paper) of the same test object have implicit local correlations that can be learned from training examples by deep neural networks. We further explore the possibility of generating phase diverse measurement(s) using a single typical measurement as may be performed by a standard imaging system. This additional pseudo-data thus generated is shown to reduce ill-posedness of the reconstruction problem, producing stable and high-quality image solutions. In what follows, we first provide the contextual background of inverse imaging problems, and subsequently describe our proposed contribution.}

The field of inverse problems in imaging has witnessed remarkable advancements leading to direct as well as iterative approaches to image reconstruction. Direct algorithms such as the Wiener filter based deconvolution remain popular due to their computational efficiency and reliance on realistic statistical assumptions about signal and noise \cite{Wiener1949}. The direct methods are computationally fast and effective for well-understood, linear problems.  For more challenging imaging problems, classical approach formulates reconstruction as an optimization problem, constituting data enforcing and regularization terms. The latter incorporates prior knowledge — such as sparsity in wavelet or gradient domains — into the reconstruction process using penalties like the $\ell_1$-norm  \cite{candes2008enhancing}, total variation \cite{rudin1992nonlinear}. The relevant optimization problems are typically solved via iterative algorithms including Landweber iteration, forward-backward splitting, ADMM, primal-dual methods or other allied approaches \cite{combettes_pesquet_2011, boyd_et_al_2011}. 

In recent years, deep learning has emerged as a new paradigm - capturing complex nonlinearities and implicit data-driven priors that classical methods struggle to model \cite{ongie2020deep}. In this respect, neural networks can be incorporated at multiple stages of imaging and inverse-problem workflows \cite{luo2025revolutionizing,lam2024}. Some methods employ neural networks for preprocessing measurements to improve the inputs to traditional reconstruction algorithms - UNet architectures, in particular, excel as preprocessors or denoisers by extracting spatial context while preserving fine detail \cite{ronneberger2015u, zhang2020holo}. Other approaches use networks in an end-to-end fashion, mapping sensor data directly to image reconstructions. This includes models such as UNet and generative adversarial networks (GANs) for tasks like Fourier phase retrieval \cite{zhang2021phasegan}. A further class of techniques integrates physical modeling into the learning process - Physics-Informed Neural Networks embed explicit models of phenomena such as diffraction or sensor response, making them well-suited for inverse problems with limited or ill-posed data  \cite{chen2020physics}. Yet another hybrid paradigm is plug‑and‑play (PnP) methods: alternating between enforcing data consistency via the physical forward model and applying a learned denoiser (often UNet or CNN), effectively combining data-driven priors with physics-based reconstruction \cite{shi2020deep}. Finally, in post‑processing approaches, neural networks (e.g. UNets or GANs) refine the outputs of classical algorithms by removing artifacts, improving resolution, or enhancing perceptual and diagnostic quality \cite{liu2019deep}. Collectively, these methodological approaches are reshaping the landscape of computational imaging. Computational imaging via deep neural networks offers the practical advantage that once the inverse mapping is learned, the reconstruction process in a field deployed imaging system can be made considerably faster compared to the conventional iterative reconstruction framework. We emphasize however that the deep neural network approach is still aiming to solve the same underlying ill-posed inverse problem as done by the conventional optimization techniques.
\begin{figure}
    \centering   \includegraphics[width=1.0\linewidth]{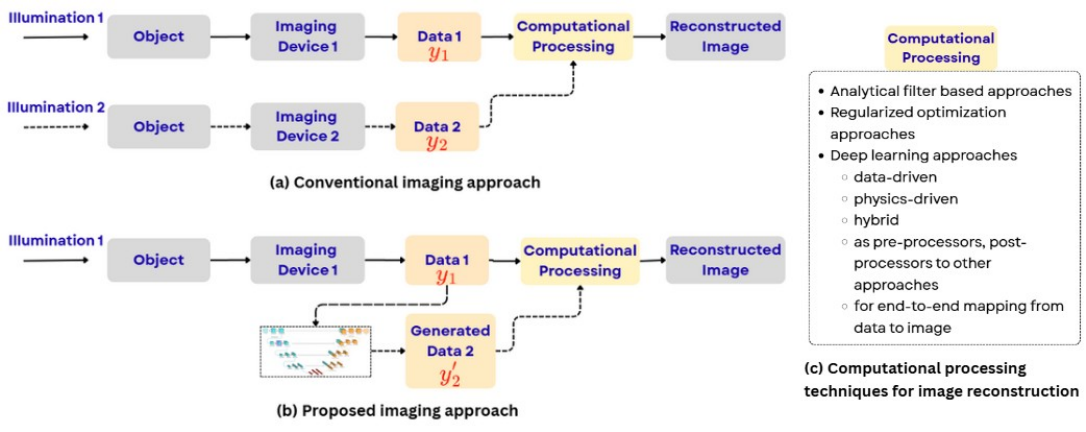}
    \caption{Imaging workflow. (a) Conventional multi-shot method acquires multiple datasets with varied illuminations/optical components for computational image reconstruction of the target object; (b) proposed method generates second-shot data using deep learning, bypassing additional hardware modifications; (c) computational processing techniques to solve imaging inverse problem.}
    \label{fig:proposed_app}
\end{figure}

It is well understood that information diversity can substantially improve the reconstruction performance. A common strategy is data augmentation, which synthetically enlarges training datasets via transformations (e.g., rotation, noise) to improve model robustness. When coupled with contrastive learning, a powerful image representation learning framework, this strategy enables models to learn invariant representations from unlabeled or sparsely labeled data leading to improved neural network performance in various computational imaging tasks \cite{le2020contrastive}. However, augmentation-based approaches may not capture complex spatial dependencies evident in real-world imaging. 
A richer alternative is to acquire physically diverse measurements via hardware modifications. 
These include multimodal acquisition, which acquire data from different sources, and multi-shot imaging, which provides multiple independent measurements of the same object within the same imaging modality. In the context of incoherent optical imaging, there are several examples where multiple measurements of the same object have provided images with superior properties. For example, the TOMBO system aims to record multiple images of the same scene via a lenslet array to lead to a thin optical imager \cite{tanida2001thin}. In this case, the recorded image scene behind each of the sub-aperture lenses is highly under-sampled, however, the composite image has superior full-pixel resolution. Structured illumination super-resolution, a microscopy modality, is another example which records multiple images of the same test object by varying the illumination in order to yield an image beating the diffraction limited resolution \cite{gustafsson2000surpassing}.
In the context of coherent imaging, a similar trend exists where the phase retrieval problem can be  
made easier through phase diversity approaches such as defocus diversity \cite{gonsalves1982phase, almoro2006complete}, spiral phase diversity \cite{sharma2015phase}, ptychography \cite{zheng2021concept, rodenburg2019ptychography}, etc.
The diversity of measured data provided by these hardware-informed methods enrich the measurement space, reduces ill-posedness, and yield more stable and accurate reconstructions.

Despite the promise of such multi-shot diverse acquisition schemes, their practical implementation suffers from increased system complexity, cost owing to additional hardware needed and requirement for careful calibration. Moreover, in many imaging systems, such as coherent electron and X-ray modalities, it is not even feasible to acquire multiple exposures as a single shot short pulse exposure may already degrade or destroy the target sample. These issues may pose challenges for widespread adoption of the multi-shot techniques although they are expected to provide a stable inverse solution. 

In this work, we introduce a new paradigm for inverse problems in imaging by fundamentally rethinking the role of data acquisition and algorithms. Rather than relying on complex multi-shot physical measurements for improving robustness of the reconstruction algorithms, we generate physics-informed pseudo-measurements through deep learning models - an innovative strategy that, to our knowledge, remains unexplored. As illustrated in figure~\ref{fig:proposed_app}, two measurements on the same object $x$ by two physical systems may produce two data $y_1 = A_1(x) + n_1$ and $y_2 = A_2(x) + n_2$, which may be used by a suitable reconstruction algorithm to produce a reconstructed image. If the measurements $y_1$ and $y_2$ provide sufficient diversity, the inverse problem solution is expected to be more robust. It is important to note that $y_1$ and $y_2$ represent measurements for the same object $x$. If the hardware systems corresponding to forward operators $A_1$ and $A_2$ provide non-redundant but implicitly equivalent information, it is reasonable to expect that the raw data vectors $y_1$ and $y_2$ have some spatial correlation. If this implicit correlation between $y_1$ and $y_2$ can be learned, it may be possible to generate a pseudo-data $y'_2$ based on the measurement $y_1$ by leveraging the abilities of deep neural networks. 

As a compelling proof of concept, we demonstrate this framework in incoherent as well as coherent optical imaging, harnessing the vortex phase diversity to enhance reconstruction quality in both imaging regimes with our simulation study. Specifically, for incoherent imaging, we show that images recorded using a 4F system with open Fourier plane aperture can be used to generate pseudo-data corresponding to images recorded with a charge-1 vortex aperture. On the other hand, in the context of coherent imaging (or phase retrieval), we show that diffraction irradiance pattern due to a plane wave illumination can be used to generate pseudo-data corresponding to the diffraction pattern for the same object but with charge-1 vortex phase illumination. In both the incoherent and coherent cases, we find that the the pseudo-data generated by deep neural network based on prior training can provide improved reconstructed images without the usual difficulties due to ill-posedness and/or stagnation problems. Our proposed framework presents an elegant solution to the limitations inherent in conventional multi-shot acquisition methods, while leveraging recent advances in deep learning and optimization-based inverse imaging algorithms.  

The paper is organized as follows. In Section 2, we first detail the prior art and challenges faced in incoherent and coherent imaging. To address these issues, in Section 3, we describe our proposed formulation based on pseudo-data generation for a robust solution to the corresponding inverse problems. This is followed by the demonstration and discussion of the results in Section 4. We finally provide the concluding remarks in Section 5.

\section{Background: related prior art and challenges}

\subsection{Incoherent imaging: PSF diversity for enhanced image contrast}

Optical imaging systems underpin a wide array of modern technologies, including mobile photography, digital pathology, defense, and surveillance applications \cite{Zhou2019, Rivenson2019}. These systems typically operate in the incoherent imaging regime, where spatially incoherent illumination is employed. The recorded image, $y_1 (u,v)$, is modeled as the convolution of the object’s true intensity, $I_{obj} (u,v)$, with the system’s point spread function (PSF), $|h(u,v)|^2$ along with detection noise. PSF of a system is dependent on the pupil/aperture function $P(f_u,f_v)$ by the Fourier relationship \cite{goodman2005introduction}:
\begin{equation}
    |h(u,v)|^2 = \bigl|\mathcal{F}\{P(f_u,f_v)\}\bigr|^2,
    \label{eq:psf}
\end{equation}
where $\mathcal{F}$ denotes the inverse Fourier transform, $(u,v)$ and $(f_u,f_v)$ are the spatial and frequency domain coordinates, respectively. We have omitted the unimportant normalization factors in the above relation. For incoherent imaging, the image formation process is linear with respect to the object intensity, corresponding to the following forward model:
\begin{equation}
	y_1(u,v) = I_{obj}(u,v) \, * \, |h(u,v)|^2 \, + n(u,v),
    \label{eq:blurring_eq}
\end{equation}
where $*$ denotes the convolution and  $n(u,v)$ represents noise in the measurement process. This convolution gives rise to image blurring, which is an inherent challenge in incoherent imaging systems. Factors such as finite aperture size, optical aberrations, and diffraction exacerbate image blurring, consequently degrading image quality and interpretability, and thus impairing performance in critical applications \cite{Rivenson2019}.
The problem of recovering an estimate of the true, sharp object from the blurred measurements, often referred to as deblurring, is inherently ill‑posed, thus requiring stabilization via regularization or data-driven constraints. 

Classical methods like Wiener deconvolution assume known noise and object power spectrum to balance the fidelity of recovered image with stability, whereas regularized deconvolution (e.g., Tikhonov, total variation) adds penalty terms that encode prior knowledge (e.g., smoothness or edges) to stabilize the solution \cite{wang2014recent}. More recently, deep‑learning models have leveraged large datasets to learn mappings from blurred to sharp images \cite{zhang2022deep}. 
\begin{figure}
    \centering
    \includegraphics[width=0.95\linewidth]{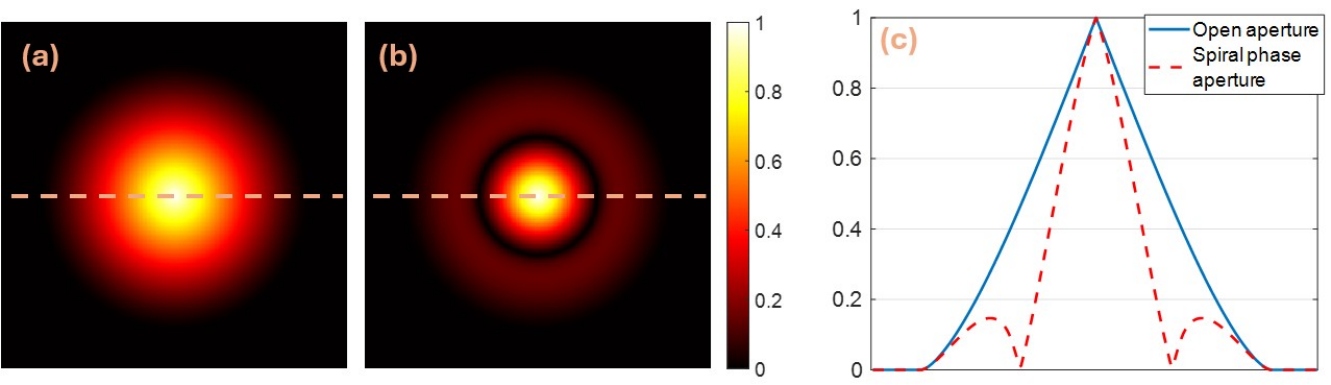} 
\caption{OTF magnitude for (a) open circular aperture and (b) spiral phase aperture, (c) profile plots of the two OTF
magnitudes along the dotted central line in (a) and (b).}
    \label{fig:otfs_mag}
\end{figure}
Deblurring techniques in incoherent optical systems also include multi-shot imaging methods, which exploit measurement diversity by acquiring multiple images under varying conditions (e.g., different illumination patterns or Fourier plane phase modulations) \cite{tahara2022review, desai2025multiple}.  This information diversity can be utilized to enhance reconstruction quality, resolution, and artifact suppression compared to single-shot approaches. Such strategies can also effectively address limitations imposed by finite apertures and diffraction, where the spatial resolution of an optical system is fundamentally restricted by its aperture size, setting a cutoff for the highest spatial frequencies recoverable due to diffraction. Recent studies have demonstrated that the high spatial frequency response can be further improved using aperture phase diversity, achieved through different aperture designs \cite{vicidomini2018sted, singh2024high}. Notably, the use of open apertures and spiral phase apertures, where charge-1 vortex phase plate can be placed in the Fourier plane of the imaging
system\footnote{For an aperture with radius $\rho_0$, the aperture function is 0 outside the radius and for $u^2 + v^2 \leq \rho_0^2 $: $P_{\text{open}}(u, v) = 1$  for open aperture and  $P_{\text{sp}}(u, v) = \exp\left[i\, \arctan\left(\frac{v}{u}\right)\right]$ for spiral phase aperture. }, modulates the pupil function. Consequently, the PSF and its frequency-domain equivalent, the optical transfer function (OTF), are tailored to enhance image reconstruction quality.
\begin{figure}
    \centering
\includegraphics[width=1.01\linewidth]{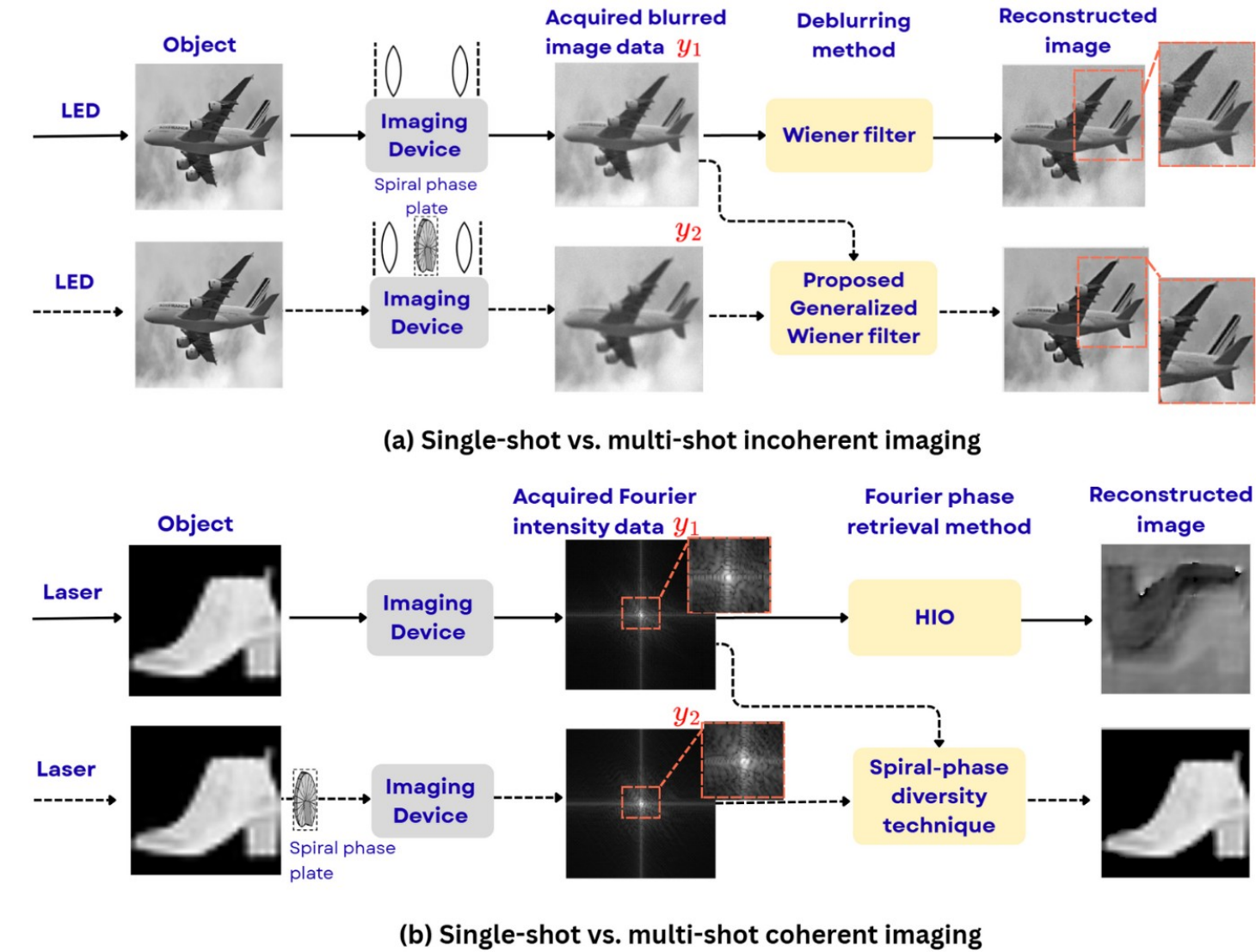}
    \caption{Single shot vs. multi-shot imaging workflow for (a) incoherent and (b) coherent regimes. For multi-shot imaging, two image data were acquired without and with a charge-1 vortex phase plate placed in Fourier plane aperture for incoherent imaging and in illumination plane for coherent imaging. }
    \label{fig:imaging_regimes}
\end{figure}
More specifically, the OTF, defined as the normalized 2D Fourier transform of the PSF, characterizes the system’s frequency response. As shown in figure~\ref{fig:otfs_mag}, an open aperture OTF decreases almost linearly with spatial frequency, while a vortex phase aperture OTF drops to zero at an intermediate frequency before rising at higher frequencies. Given the same aperture radius, both the OTFs have same cut-off frequency but their frequency response is distinct. By combining measurements from both aperture types, the diverse OTFs can provide enhanced sensitivity to high spatial frequencies, thus improving image contrast and detail. Using both the open aperture and the vortex aperture in a linear imaging configuration, the imaging process~\eqref{eq:blurring_eq} for each aperture configuration is described as:
\begin{equation}
	y_k(u,v) = I_{obj}(u,v) \, * \, |h_k(u,v)|^2 \, + n_k(u,v),
    \label{eq:blurring_eq2}
\end{equation}
and in the frequency domain, it may be represented as
\begin{equation}
	\tilde{y}_k (f_u, f_v) = \tilde{I}_{obj} (f_u, f_v) \, OTF_k(f_u, f_v) \, + \, N_k (f_u, f_v),
    \label{eq:blur_model}
\end{equation}
where $k=(1,2)$ corresponds to the terms with open and vortex aperture, respectively, $\tilde{I}_{obj}$ is the object’s spectrum, and $N_k$ represents noise in the Fourier domain. This framework, depicted visually in figure~\ref{fig:imaging_regimes}(a), enables the retrieval of richer spatial information by leveraging optical hardware diversity. To optimally combine images captured through different apertures, Generalized Wiener (GW) filter extends classical Wiener filtering to the multi-shot data context. Basically, for acquired images $y_k$, we seek a frequency domain filter $W_k$ which minimizes the following expected mean squared error:
\begin{equation}
    \epsilon =  \langle ||\tilde{I}_{obj} - \sum_k W_k \, \tilde{y}_k ||^2 \rangle .
\end{equation}
Minimizing the above expression yields the following expressions for GW filters:
\begin{equation}
	W_k(f_u,f_v) = \frac{OTF_k^*(f_u,f_v)}{\sum_k |OTF_k(f_u,f_v)|^2 + \frac{S_n(f_u,f_v)}{S_{obj}(f_u,f_v)}} ,
    \label{eq:GW_filter}
\end{equation}
with $S_{obj}$ and $S_n$ representing the object and noise spectral density, respectively.
Depicted as standard approach in figure~\ref{fig:prop_techniques}(a), the true object can then be estimated using
\begin{equation}
    \hat{I}_{obj} = \mathcal{F}^{-1} \Bigl\{\sum_k W_k \, \tilde{y}_k \Bigr\}.
\end{equation}
It is important to note that the OTFs corresponding to the native system aperture and an aperture with additional vortex phase do not generally share common zeros. This characteristic ensures that the GW filter is not hindered by missing frequency components as is the case for single-shot deblurring with the Wiener filter. As shown in figure~\ref{fig:imaging_regimes}(a), the reconstructed image obtained by GW filter is superior to Wiener filtered image which shows faint grainy artifacts. 

Although this approach shows promise for improving imaging contrast, it requires two acquisitions per image—a requirement that may introduce practical difficulties. In section~\ref{sec:prop_form}, we describe our contribution where we generate the data corresponding to $y_2$ with a deep neural network, thus offering the possibility of enhanced imaging performance without physically recording the image data with vortex phase in the Fourier plane.

\subsection{Coherent imaging: Phase diversity for resolving stagnation of phase retrieval algorithms}
Coherent imaging modalities, such as coherent X-ray diffractive imaging, electron microscopy, and wavefront sensing, are integral to advanced imaging techniques. These systems involve the illumination of the object with a coherent plane wave. The resulting diffracted field is recorded on a detector in the far-field regime, which corresponds to the two-dimensional Fourier transform of the object function. 
For a complex valued object field $\bar{I}_{obj} (u,v)$, the current detectors only measure the intensity, $y_1 (f_u, f_v)$ of the Fourier transformed field leading to the following forward model: 
\begin{equation}
    y_1 (f_u, f_v) = |\mathcal{F} \{\bar{I}_{obj} (u,v)\} |^2.
    \label{eq:coherent_im}
\end{equation}
The objective of the corresponding inverse problem is to reconstruct the complex object field $\bar{I}_{obj} (u,v)$ solely from the Fourier intensity data. This problem, commonly known as Fourier phase retrieval, is a severely ill-posed due to the loss of phase information, leading to multiple possible solutions with same intensity measurement. Among the most persistent challenges in coherent Fourier phase retrieval is the twin-image problem \cite{guizar2012understanding_twin_image_problem}: an object and its inverted complex conjugated version share the same Fourier magnitude. As a result, standard iterative algorithms, such as Gerchberg–Saxton (GS) and hybrid input–output (HIO), frequently oscillate between these equivalent solutions without achieving stable convergence. Moreover, these approaches are computationally demanding due to their sensitivity to initialization. Multiple runs with different random initializations are typically required, with plausible ``good quality" reconstructions subsequently selected based on some metric and then averaged with phase adjustment to get a solution estimate.
\begin{figure}
\includegraphics[width=1.01\linewidth]{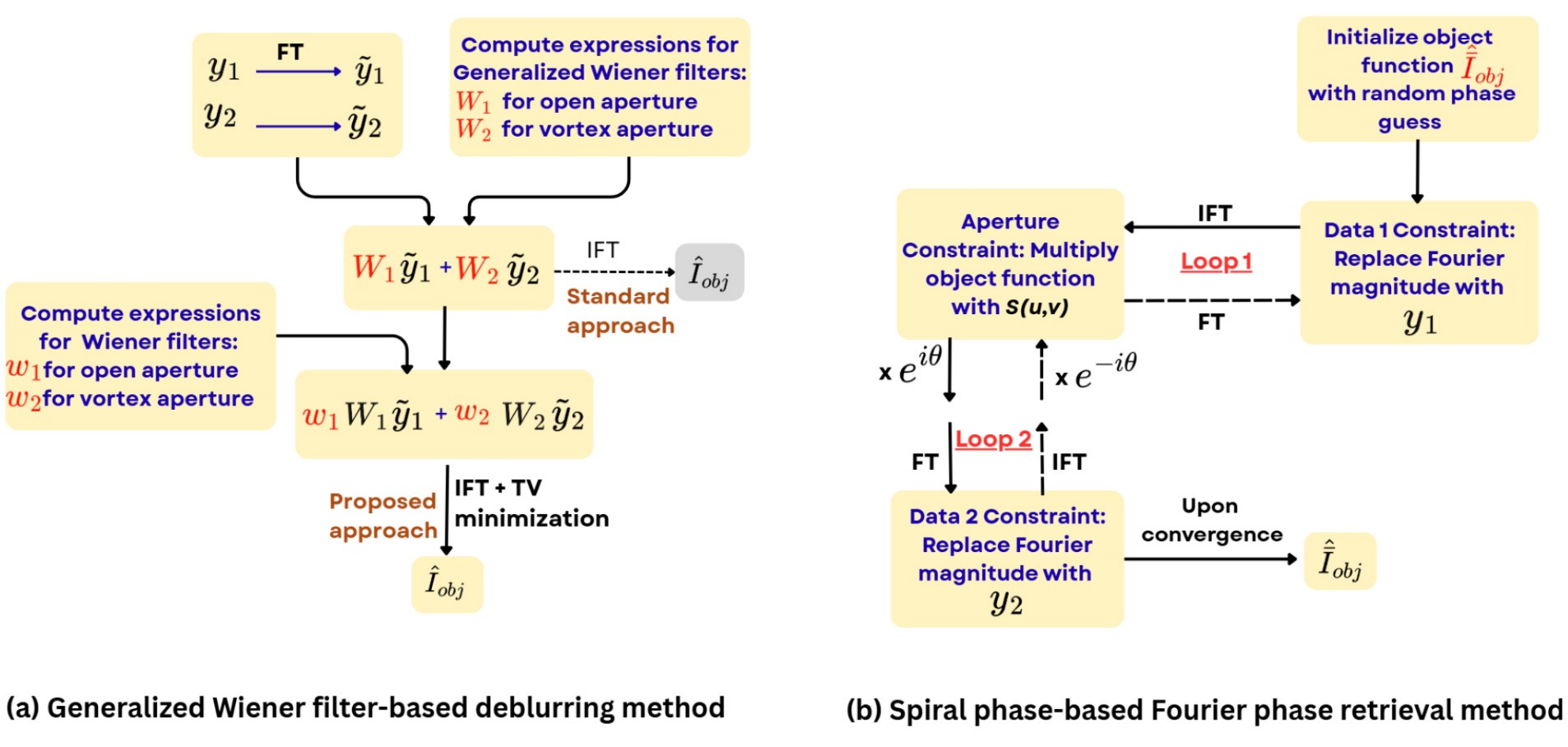}
     \caption{Computational techniques used to reconstruct images from the multi-shot data for (a) incoherent imaging and (b) coherent imaging case. FT stands for Fourier transform, IFT for inverse Fourier transform.}
     \label{fig:prop_techniques}
\end{figure}

Few works have been reported in the literature to combat the issue of twin stagnation. This includes Complexity-Guided Phase Retrieval (CGPR) that estimates a “complexity” metric from the Fourier magnitude and enforces sparsity in a controlled manner in the iterative solution \cite{butola2019CGPR}. While helping the conventional phase retrieval algorithms escape stagnation and reducing artifacts, CGPR still demands heavy computation and lacks finite-iteration convergence guarantees. More recently, \cite{kularia2024twin} demonstrated that single-shot charge-1 vortex illumination-based diffraction intensity measurement can inherently resolve the twin-image ambiguity. The vortex beam introduces an isolated zero near the DC frequency in the diffraction data. The iterative Fourier transform  algorithms, like the HIO method, tend to develop either a clockwise or anti-clockwise vortex phase structure centered at the amplitude zero, effectively biasing early iterations of HIO towards one of the twin solutions. This deterministic bias leads to consistent convergence to a twin-free reconstruction within a few hundred iterations. Deep learning approaches to Fourier phase retrieval have also been explored in various configurations to enhance reconstruction accuracy and efficiency \cite{lam2024,dong2023phase}.

Another widely adopted strategy to alleviate twin-image and other stagnation artifacts in phase retrieval is the use of multi-shot phase diversity methods, which exploit measurement diversity to promote solution uniqueness. Unlike single-shot approaches, multi-shot techniques can achieve convergence within substantially fewer iterations by leveraging the additional information encoded in multiple, diverse observations. For instance, \cite{sharma2015phase} addressed the phase retrieval problem by exploiting the diversity of diffraction patterns generated with both plane-wave and vortex-beam illumination, and providing a twin-stagnation free image reconstruction. As illustrated in figure~\ref{fig:imaging_regimes}(b), image reconstructed using a standard approach like HIO often suffers from twin-image stagnation, whereas the multi-shot approach eliminates this artifact. Mathematically, the imaging process~\eqref{eq:coherent_im} for each aperture configuration translates to:
\begin{align}
y_1 (f_u, f_v) &= | \mathcal{F} \{ \bar{I}_{obj} (u,v) \, S(u,v) \}|^2, \nonumber \\
y_2 (f_u, f_v) &= | \mathcal{F} \{ \bar{I}_{obj} (u,v) \, S(u,v) \, e^{\iota \theta(u,v)} \}|^2,
 \label{eq:fourier_intensity}
\end{align}
\noindent where \( S(u, v) \) is the lens aperture/support function. $y_1$ and $y_2$ corresponds to the measured Fourier intensity using plane wave and vortex beam illumination, respectively. In the latter case, $\theta (u,v) = \tan^{-1} (v/u)$ represents the spiral phase term. 
The iterative algorithm, based on the GS framework, alternates between the two measurement planes. As depicted in figure~\ref{fig:prop_techniques}(b), the field corresponding to the first measurement is initialized in the Fourier domain with a random phase and the amplitude as the square root of the measured Fourier intensity $y_1$. This field is then inverse Fourier transformed to the object domain with the aperture constraint applied (i.e., finite support constraint), and modulated by a spiral-phase function $e^{i \theta}$ before propagating via Fourier transform to the second measurement plane. There, the amplitude is updated as per the measured intensity $y_2$ while preserving the phase. This is followed by the backpropagation via inverse Fourier transform of the estimated field with conjugate spiral phase demodulation and re-application of the aperture constraint. This iterative enforcement of Fourier magnitude and aperture constraints across both planes is repeated until a stable phase solution is obtained. 

Despite its effectiveness in suppressing stagnation artifacts, implementing spiral-phase illumination requires additional hardware and is often impractical—especially in shorter-wavelength regimes like X-rays or electron beams. As a result, most Fourier phase retrieval systems in these domains still rely on conventional plane-wave illumination. In the next section, we present our approach to overcome this limitation via the capabilities offered by deep learning.

We wish to emphasize that in both incoherent and coherent imaging cases discussed in this section, it is clear that availability of the second data $y_2$ makes it considerably easier to solve the corresponding inverse problem. Generating an approximation to $y_2$ therefore appears to be an attractive alternative to solving an inherently ill-posed problem through a single measurement $y_1$. 
We discuss the details of how deep learning can effectively help us achieve this goal in the next section.

\section{Proposed formulation}
\label{sec:prop_form}

In this work, we augment single-shot imaging by generating a synthetic ``second shot" through deep learning, eliminating the need for physical dual measurements for both incoherent and coherent imaging cases described in earlier discussion. To be more precise, we train a neural network on simulated paired data ($y_1, y_2$) to learn the implicit spatial relationship between the first data $y_1$ to a diversity-enhancing data $y_2$. Once trained, the neural network model can provide an approximation ($y'_2$) of $y_2$ from the single-shot measurement ($y_1$). Our task is to understand if the use of $y_1$ and $y'_2$ can lead to a good quality inverse solution without performing actual physical measurement $y_2$. We observe that the data in pair ($y_1$, $y_2$) have the same dimensions and are expected to have a local spatial correlation for both incoherent and coherent imaging cases when charge-1 vortex phase diversity is employed. The relation between $y_1$ and $y_2$ is therefore an image-to-image map. {We therefore choose the well-established UNet architecture for learning this relationship from training data set for both incoherent and coherent imaging case.}

\subsection{Network architecture}

UNet is a fully convolutional neural network recognized for its “U”-shaped architecture, designed for image-to-image translation tasks \cite{ronneberger2015u}. It features a contracting encoder to capture global context and an expanding decoder to restore local details, with shortcut skip connections between encoder and decoder to preserve spatial information.
\begin{figure}
    \centering
\includegraphics[scale=0.48]{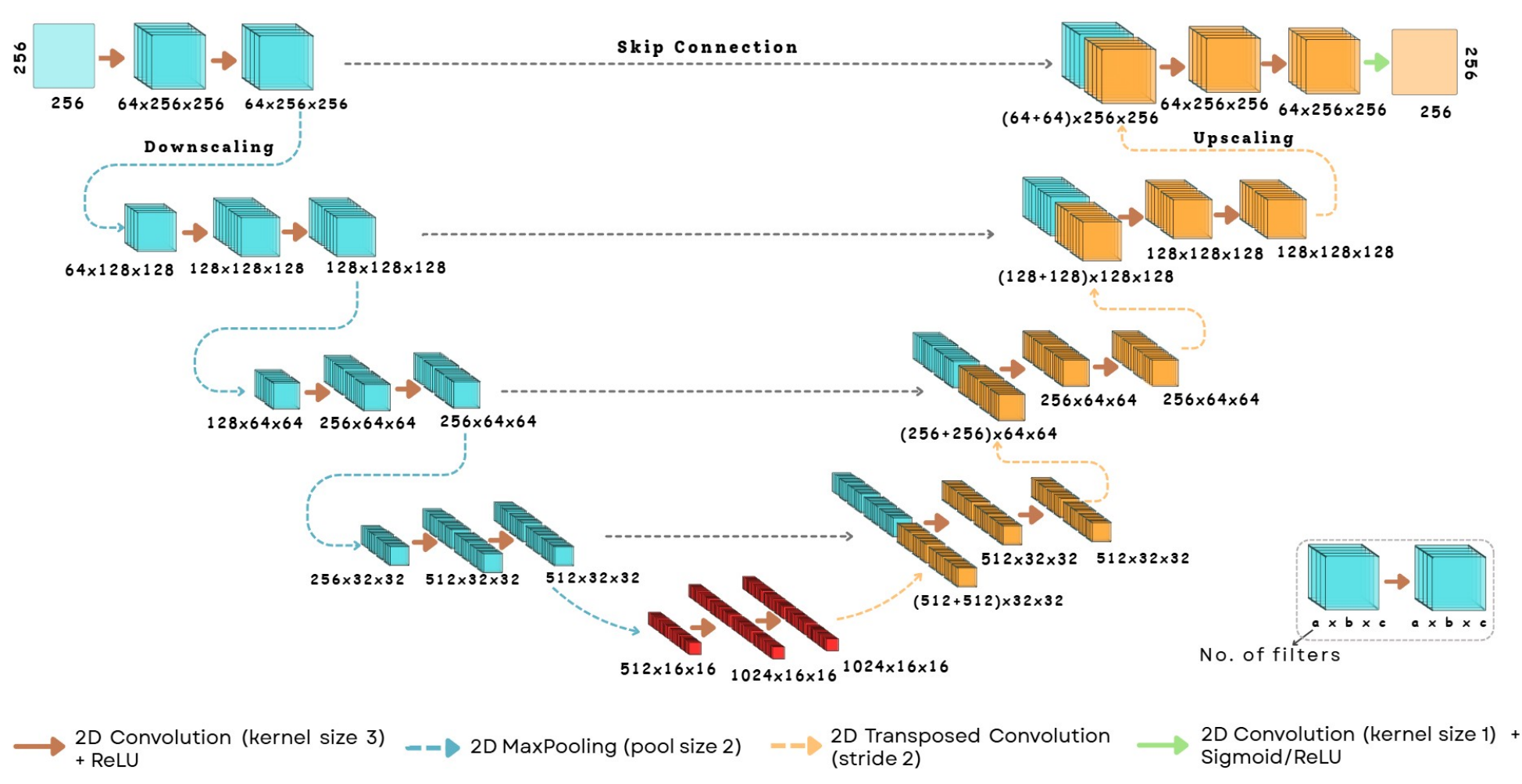}
    \caption{UNet based deep learning architecture used to generate pseudo-data for both coherent and incoherent imaging.}
    \label{fig:unet}
\end{figure}
As depicted in figure~\ref{fig:unet}, our adopted model comprises of four encoding (in blue) and decoding (in orange) stages. Each encoder block has two consecutive 3×3 convolutions with ReLU activation followed by 2×2 max pooling. This double convolution effectively increases the receptive field and enriches structural information passed to the consecutive layers. The bottleneck block (in red) includes two convolutional layers producing a compact latent representation.
\begin{table}
\centering
\begin{tabular}{|c |c|c|c|c|}
\hline
\textbf{Imaging } & \textbf{Training/test  } & \textbf{Training } & \textbf{Testing } & \textbf{Mean SSIM}\\
\textbf{ modality} & \textbf{data domain} & \textbf{ data size} & \textbf{ data size} & \textbf{$\pm$ SD}\\
\hline
Incoherent imaging & Image domain & 1732 & 434 & $0.97 \pm 0.01$\\
\hline
Coherent imaging & Fourier domain & 6000 & 1000 & $0.89 \pm 0.05$ \\
\hline
\end{tabular}
\caption{{Dataset description with mean SSIM values computed on test data after network training. The training data was further divided into training and validation subsets in an 85:15 ratio.}}
\end{table}
Decoder blocks use transposed convolutions for up-sampling, concatenate encoder features via skip connections, and apply two 2×2 convolutions with ReLU. The final layer uses a 1×1 convolution and an application-specific activation function (sigmoid for intensity images in incoherent imaging case, ReLU for Fourier intensity images in coherent imaging case) to generate the output. This design enables effective transformation learning between input ($y_1$) and target ($y_2$) data pair, enhancing structural preservation and feature extraction. To train the model, we used the Adam optimizer with a learning rate of 0.001. For the loss function, we employed a hybrid term combining Mean Squared Error (MSE) and Structural Similarity Index Measure (SSIM) as follows:
\begin{equation}
        \mathcal{L} =  \text{MSE} + \alpha \, (1 - \text{SSIM}) \,,
    \end{equation}
where $\alpha$ is a weighting factor that modulates the relative contribution of the two terms. In the present study, $\alpha$ was set to 1. MSE computes the mean squared difference between the true and the predicted images, ensuring pixel-wise reconstruction fidelity by minimizing numerical errors. The SSIM metric (values ranging between 0 and 1), on the other hand, quantifies the structural similarity, ensuring structural fidelity in the reconstructed pseudo data. The lower the MSE and higher the SSIM, the better the reconstruction. For evaluation purpose, we tracked SSIM and peak signal-to-noise ratio (PSNR) over epochs during the training phase. These metrics provided insight into both numerical accuracy and structural fidelity, ensuring that the reconstructed pseudo-data approximation $y'_2$ is of desired quality to be used in the subsequent inverse solution.  It is to be noted that this work is mainly geared towards demonstrating the pseudo-data generation concept and so there is always a scope for going with alternate network architectures or loss functions.

The model was trained in PyTorch using a Nvidia RTX 3060 GPU with 6GB of VRAM. The model was trained for 50 epochs, with a training time per epoch as $\sim$ 1 minute and $\sim$ 4 minutes for incoherent and coherent imaging cases, respectively. The training and testing data sizes are summarized in Table 1. The two trained networks for incoherent and coherent cases were then used with the test data to generate $y'_2$ from $y_1$. The inference time to generate pseudo-image given the first image was less than a second ($\sim$ 0.1 seconds).

\subsection{PSF diversity for enhanced image contrast}
In this case, our adopted UNet model was trained to convert intensity images ${y}_1$, captured with an open aperture in Fourier plane, into corresponding images $y'_2$, simulating those obtained with a vortex aperture~\eqref{eq:blurring_eq2}. For this purpose, images used for training were taken from Coco128 dataset \cite{lin2015microsoft}, \cite{usc_sipi_image_database}, and resized to \( 256 \times 256 \). We defined an aperture radius of 50 pixels to compute the corresponding PSF via \eqref{eq:psf}, then convolved this PSF with each image in the dataset to generate the blurred inputs. Similarly, the PSF for the vortex aperture was used to produce a second set of blurred images, which served as the network’s training targets. The
blurred images had a 1\% Gaussian noise added synthetically.

Once the network was trained, the pseudo-generated image pairs ($y_1, y'_2$) were combined using GW filter (figure~\ref{fig:prop_techniques}(a)), which generally works well to produce the underlying sharp image. Image reconstruction performance, however, can further be enhanced by making use of cascaded filters \cite{lahmiri2017iterative, habeeb2023medical}. These filters iteratively refine the image by isolating and treating noise and blur in low- and high-frequency regions separately. Multiple Wiener filtering stages sequentially correct residual blur and noise left by previous steps, improving fine detail restoration and artifact suppression compared to a single application. In the current work, we propose a new scheme for cascading filters to deblur images from a pair of images. More precisely, we propose to apply Wiener filters after application of GW filter. Mathematically, our proposed formulation resorts to tackle image deblurring using the following equation:
\begin{equation}
    \hat{I}_{obj} = \mathcal{F}^{-1} \Bigl\{\sum_k w_k \, W_k \, \tilde{y}_k \Bigr\}.
\end{equation}
TV reduction steps can further be applied to enhance image reconstruction quality \cite{singh2024high}. The proposed scheme is highlighted in figure~\ref{fig:prop_techniques}(a). 
\begin{figure}
    \centering
\includegraphics[scale=0.6]{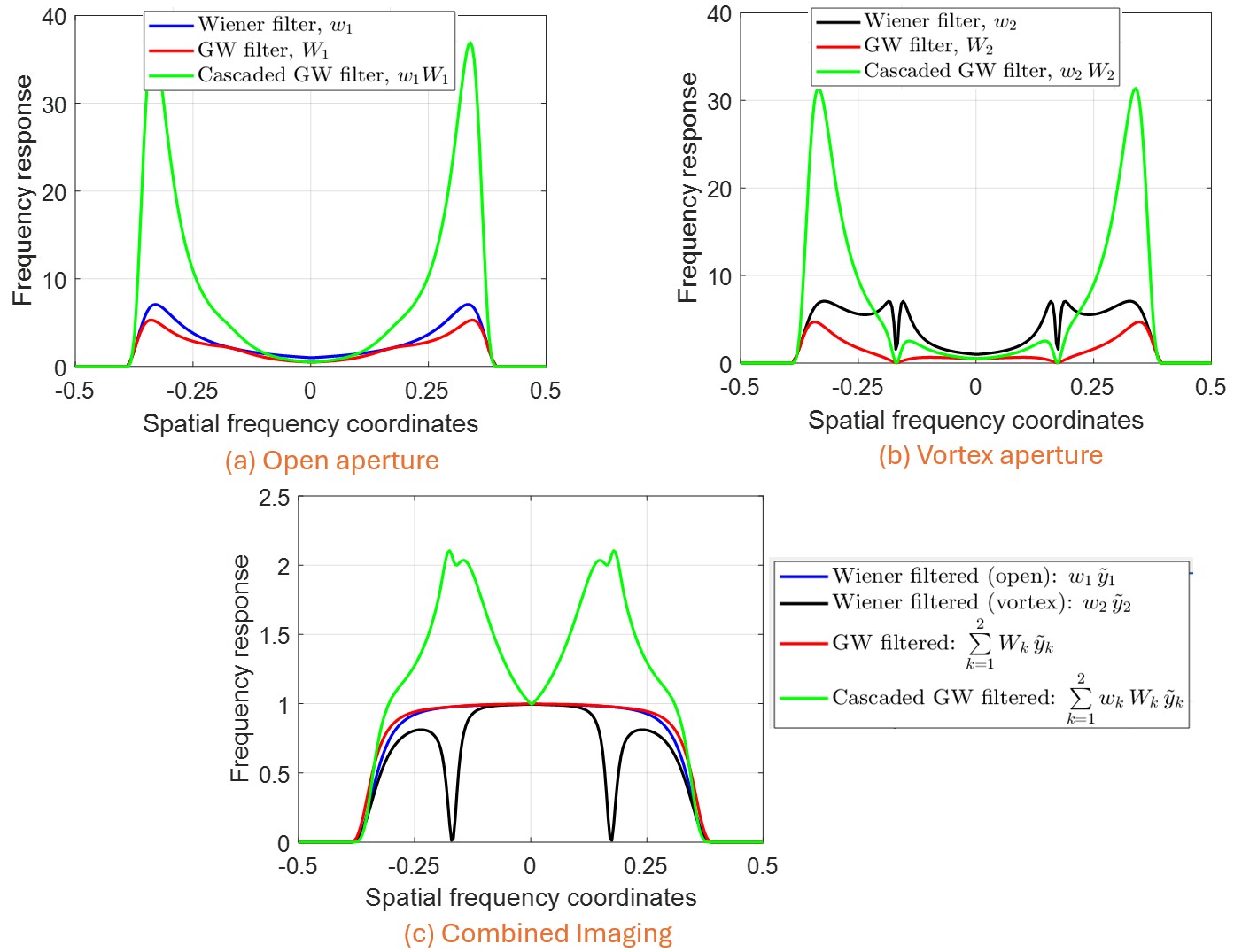}
    \caption{Central line profiles of Fourier domain magnitudes of different filters for (a) open and (b) vortex phase
apertures. (c) shows the corresponding profiles in the case of combined imaging of a point source using both apertures. These plots represent the magnitude spectra of Fourier-domain images obtained after filtering the respective acquired images from the two apertures. Cascaded GW filter can be seen to have enhanced high spatial frequency response compared to Wiener and Generalized Wiener (GW) filters.}
    \label{fig:filters_freq_resp}
\end{figure}
{Further, to highlight the advantage of the cascaded GW filter over other filters, figure~\ref{fig:filters_freq_resp} presents the Fourier domain magnitude profiles along the central line for different cases. Panels (a) and (b) show these profiles for Wiener, GW and cascaded GW filters under open aperture and vortex phase aperture conditions, respectively. For combined imaging of a centered point source, $I_{obj}$ using both apertures (as described in~\eqref{eq:blurring_eq2},~\eqref{eq:blur_model}), panel (c) depicts the resulting filter responses in Fourier domain. Specifically, it illustrates the central line magnitude-profiles of images obtained after applying these filters to the respective Fourier domain images, $\tilde{y}_1, \tilde{y}_2$, which for a point source correspond to their respective aperture OTFs (as shown in figure~\ref{fig:otfs_mag}). From these plots, it is evident that the cascaded GW filter (green curve) achieves superior enhancement of high spatial frequencies compared to the diffraction-limited responses of the individual apertures.}

\subsection{Phase diversity for resolving stagnation in phase retrieval algorithms}
In this case, the UNet model was trained to map Fourier intensity measurements due to plane wave illumination \( {y}_1 \), to their spiral-phase-modulated counterparts \( y_2 \) using paired simulated data~\eqref{eq:fourier_intensity}. Fashion MNIST dataset was used for training the network \cite{lecun2010mnist}. Using these images, pure phase objects were generated with phase values in the range \( [0, 2\pi/3] \). The overall computational window was taken to be \( 256 \times 256 \), while the object support was limited to \( 100 \times 100 \) pixels at the center to satisfy the Nyquist criterion. {The object's Fourier magnitude raised to the power 0.1 yielded \( y_1 \), while \( y_2 \) was obtained by multiplying the object with a spiral phase function prior to taking the Fourier magnitude and raising it to the power 0.1. Power of 0.1 was used to reduce the dynamic range of the diffraction data thus facilitating network training. Without this scaling, typical Fourier magnitude data exhibits large values concentrated near the center and near-zero values elsewhere. Such data provides insufficient detail for the network to learn effectively. {In practice, the data can be raised to any suitable power. In our case, preliminary results suggested using 0.1 gave good results.} 

Once the network was trained, previously explained spiral-phase diversity based method was applied to $(y_1, y'_2)$. To further refine the solution, we introduced a minor adjustment in the final stages of the algorithm. Specifically, during the last 25 iterations - when the reconstruction had already converged to a reasonably accurate estimate - we disabled the spiral-phase-modulated part of the loop and continued the iterations using only the measurement, \( y_1 \). This step improved data fidelity by reinforcing consistency with the ground truth measurement.

\section{Results and discussion}
\begin{figure}
    \centering
    \includegraphics[width=1.01\linewidth]{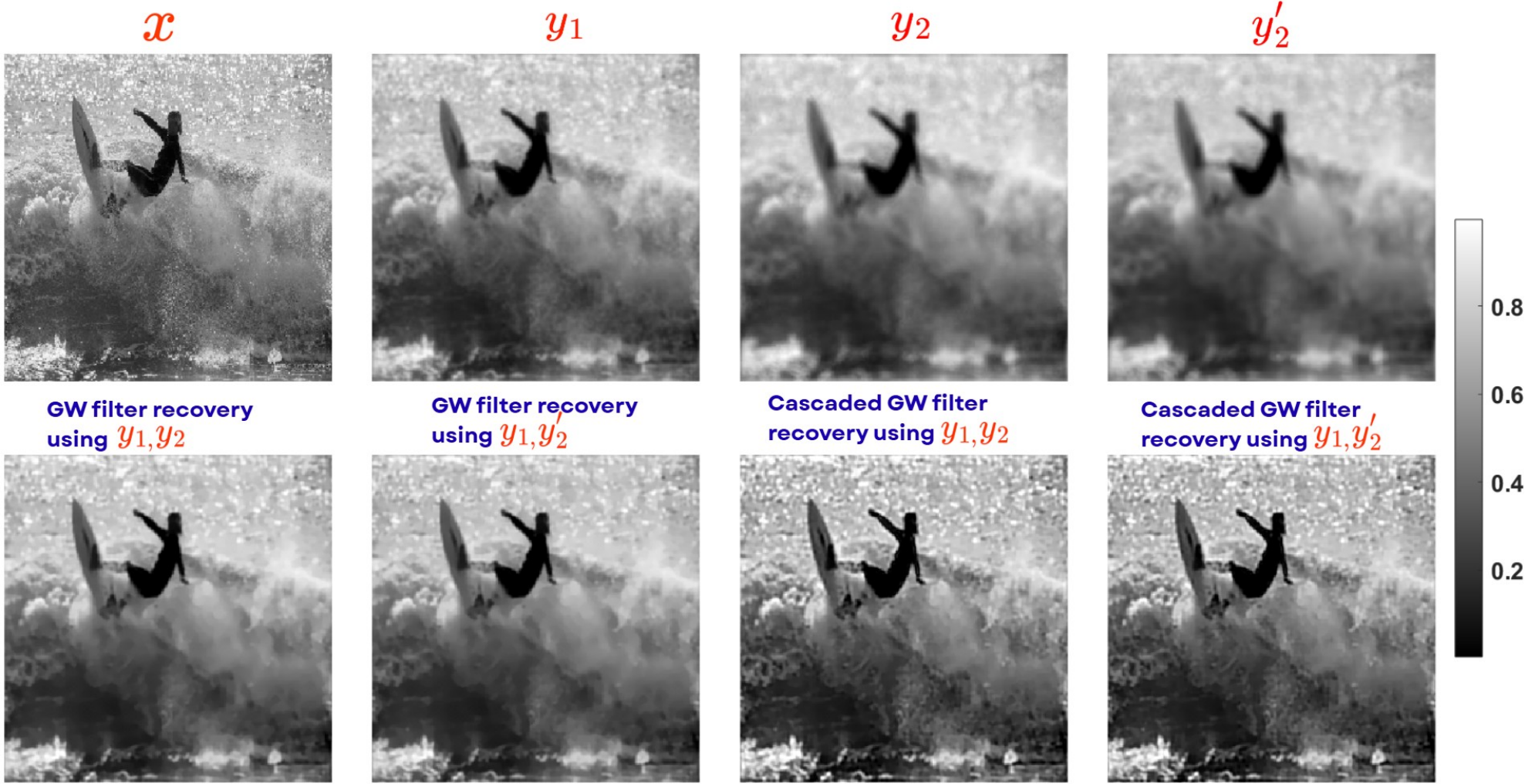} \\
    \vspace{0.2cm}
    \includegraphics[width=1.01\linewidth]{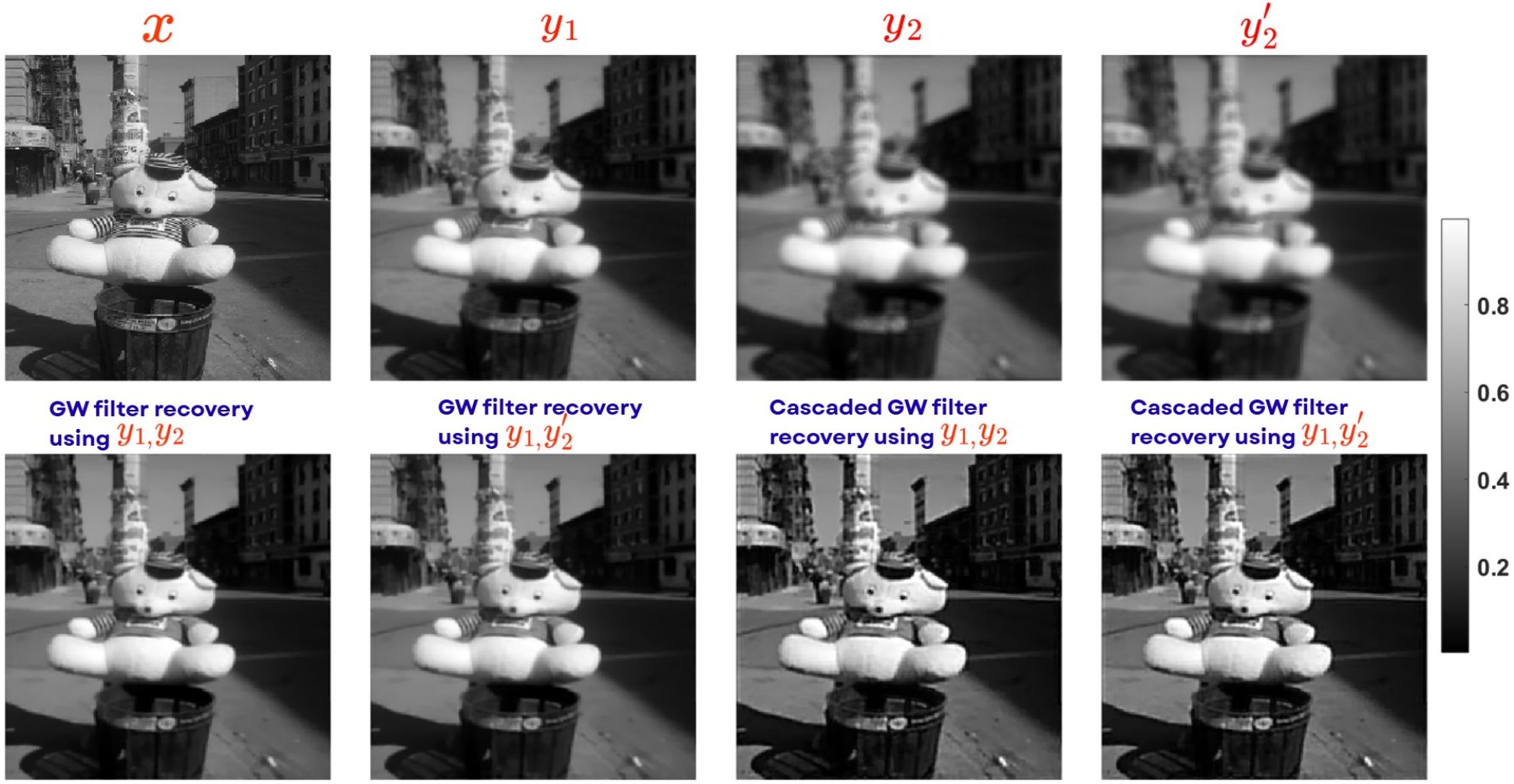}
    \caption{Comparative analysis of UNet-generated pseudo-data, $y'_2$ vs. true data, $y_2$, for two test images, $x$ (first and third rows) in incoherent imaging; $y'_2$ closely matches $y_2$. The second and fourth rows show reconstructions using the GW filter and cascaded GW filter, with the latter demonstrating enhanced spatial contrast.}
    \label{fig:images_res_deblur}
\end{figure}

\begin{figure}
    \centering
    
    \includegraphics[scale=0.5]{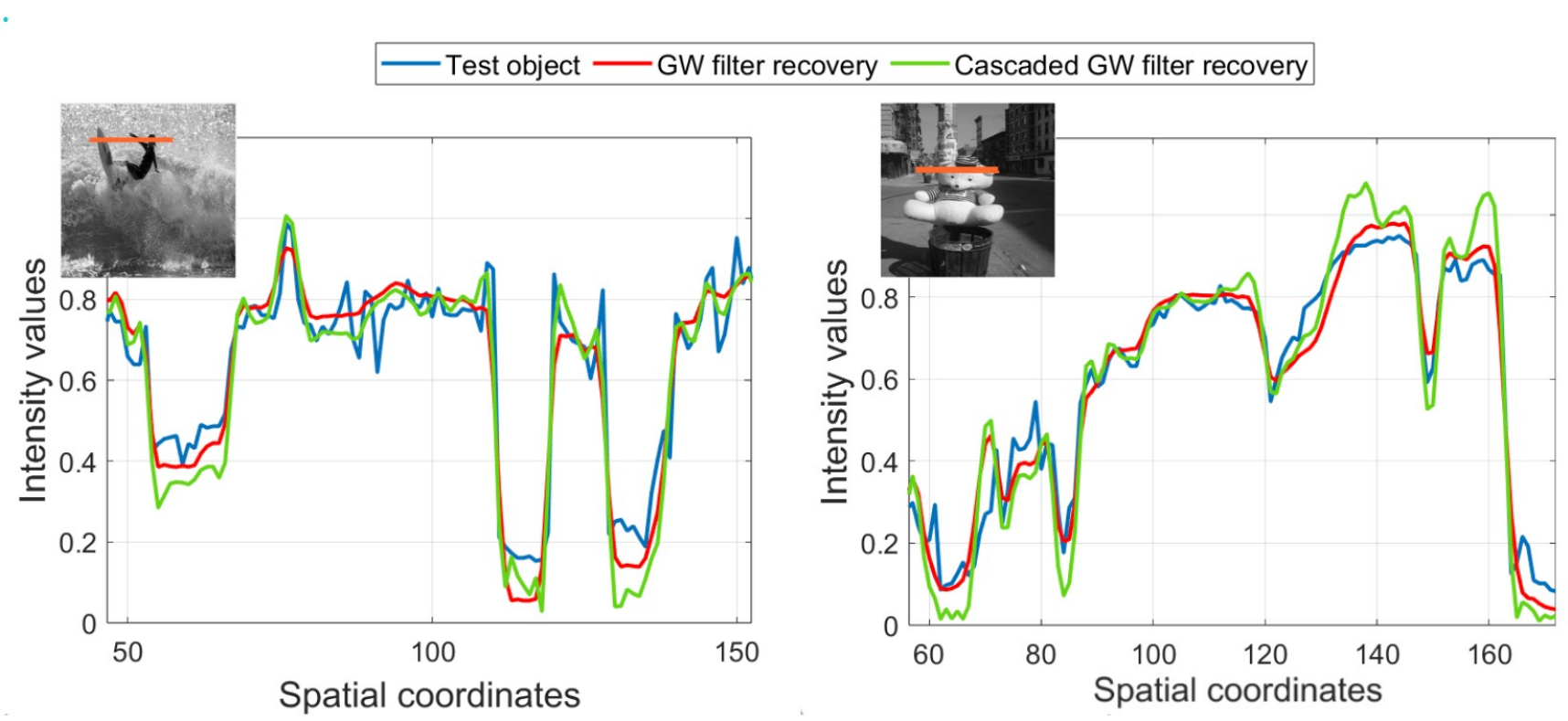}
    \caption{Line plots for the two test images in figure~\ref{fig:images_res_deblur}, along the orange line plotted on top of image inlets, compare the true object (blue line), GW filter (red line), and cascaded GW filter (green line) recoveries using $(y_1, y'_2)$. The latter shows enhanced spatial contrast.}
    \label{fig:profile_plots}
\end{figure}

\begin{figure}
    \centering
\includegraphics[width=0.7\linewidth]{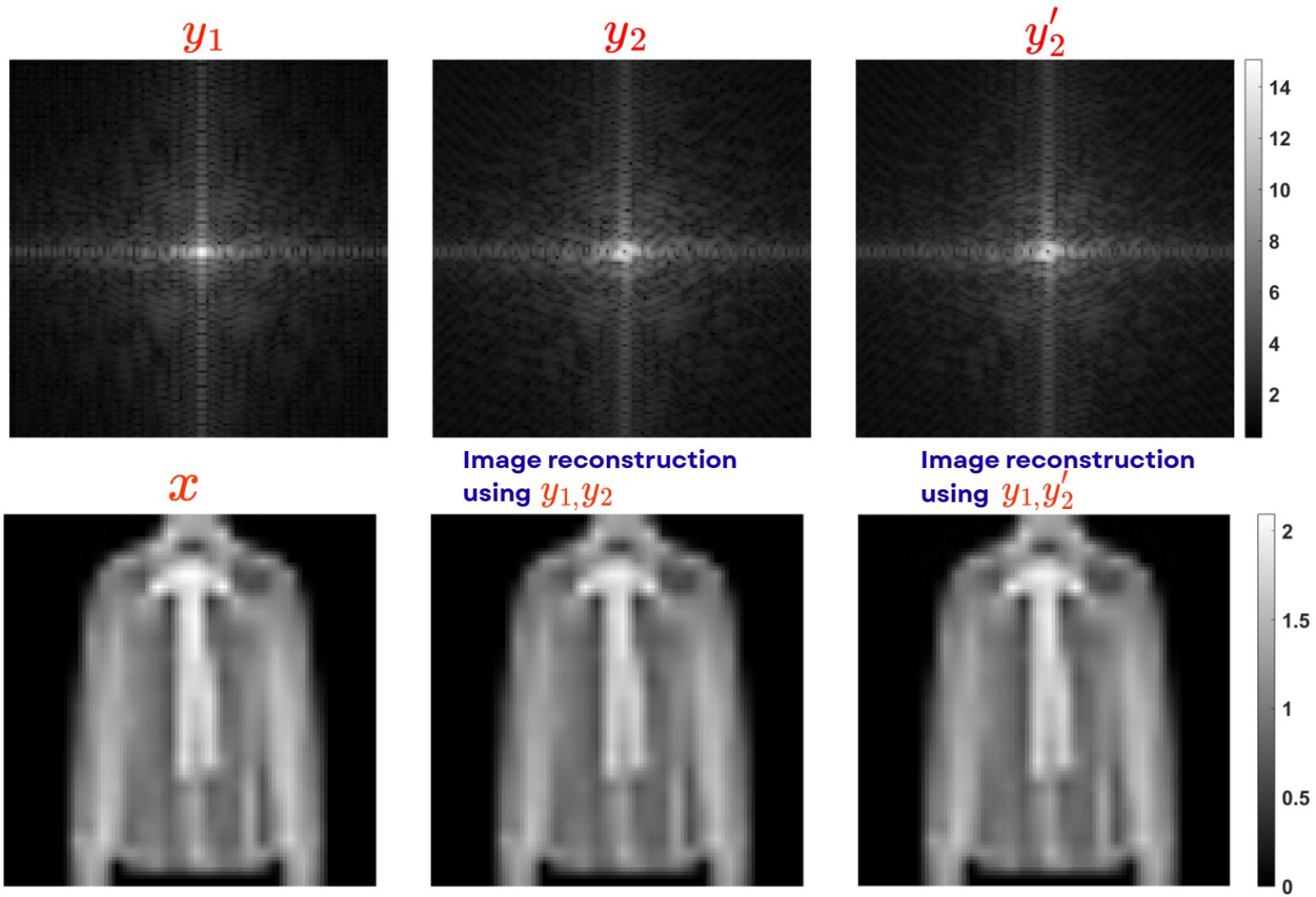} \\
 \vspace{0.1cm}
    \includegraphics[width=0.7\linewidth]{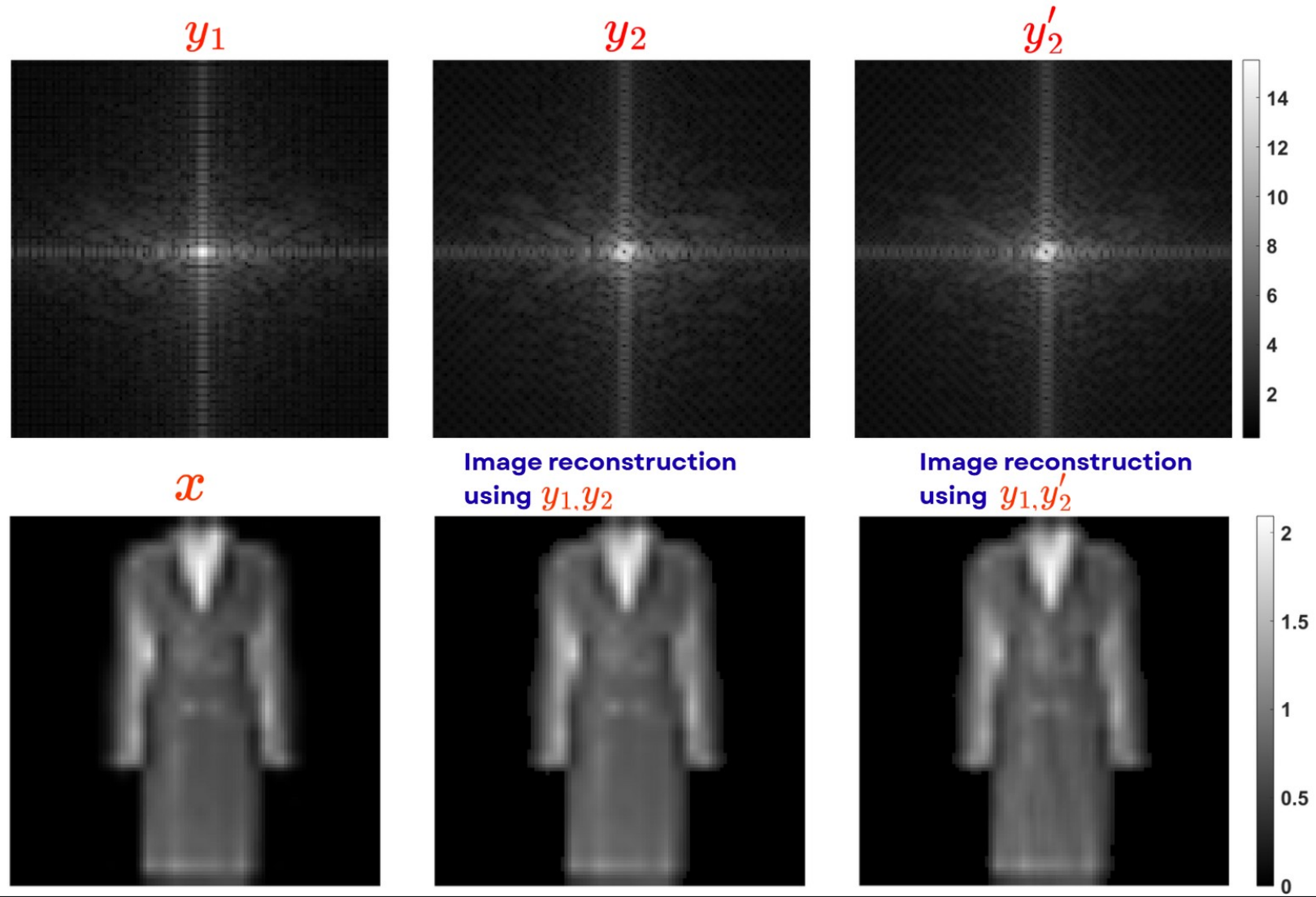} \\
    \caption{Comparative analysis of UNet-generated pseudo-data, $y'_2$ vs. true data, $y_2$, for two test images (first and third rows) in coherent imaging; $y'_2$ closely matches $y_2$. Zoomed-in Fourier magnitude images (i.e., $y_1, y_2, y'_2$) near the DC region are displayed with values raised to the power of 0.3. The second and fourth rows show the true object, $x$, reconstructions by spiral-phase based Fourier phase retrieval method using $(y_1, y_2)$ and $(y_1,y'_2)$. Only the object support region is
displayed. }
    \label{fig:images_res_fpr}
\end{figure}

To evaluate UNet's performance in generating $y'_2$ from $y_1$, we measured SSIM and PSNR between $y_2$ and $y'_2$. 
For the test set in incoherent imaging case, mean SSIM was evaluated to be $0.97 \pm 0.01$, whereas PSNR was $37.84 \pm 3.91$ dB, indicating reliable generation of pseudo-data. 
Figure~\ref{fig:images_res_deblur} visually compares two sets of images from the test data. First row in both the cases demonstrates that generated $y'_2$ closely matches $y_2$, having an SSIM of 0.98. Second row compares reconstructions via standard and proposed GW filter approaches. Both methods were applied to simulated $(y_1, y_2)$ and pseudo-generated pairs $(y_1, y'_2)$. It can be observed that even the image recoveries using $(y_1, y'_2)$ closely resemble those obtained using $(y_1, y_2)$, with SSIM of 0.99 in the examples shown. Although both generalized and cascaded GW filters provided good de-blurring performance, the cascaded version yielded improved contrast and sharper images. This is further supported by the profile plots shown in figure~\ref{fig:profile_plots}. Green lines, representing the line plot for the proposed cascaded GW approach, exhibited higher peak and lower trough values, signifying enhanced spatial contrast.

Similarly, for the coherent imaging case, mean SSIM and PSNR computed on the test set were $0.89 \, \pm  \, 0.05$ and $33.62 \, \pm \, 1.71$ dB, respectively, confirming reliable pseudo-data generation by UNet. 
Figure~\ref{fig:images_res_fpr} depicts visual comparison of the results obtained using two sets of Fourier magnitude data. In the first and third rows, the Fourier domain data are shown, where it can be seen that UNet generated $y'_2$ resembles the true data $y_2$, with an SSIM of 0.91 for the first set, 0.93 for the second set. This is followed by the image reconstruction results in the respective next rows, where spiral-phase diversity based Fourier phase retrieval was applied on both simulated data pair $(y_1, y_2)$ and generated pseudo-data pair $(y_1, y'_2)$. The two images can be observed to be similar to each other, having an SSIM of 0.97 and 0.95 in the two sets, respectively.

The results demonstrate the feasibility and reliability of generating a physically diverse second set of measurements from a single-shot measurement for both incoherent and coherent imaging cases showcasing its broad applicability to multiple optical imaging modalities. While there is a potential for using multiple diversity mechanisms, the use of spiral phase filter as a diversity mechanism demands a brief discussion. The spiral phase filter in Fourier domain is known to be the 2D analogue of the well-known 1D Hilbert transform \cite{larkin2001natural, khare2008complex} and therefore in some sense provides optimal diversity in the data. In the coherent case, this property leads to complementary diffraction patterns where the positions of maxima and minima in the diffraction intensity pattern are exchanged. {For instance, it can be observed in figure~\ref{fig:images_res_fpr} that the DC region of the vortex beam diffracted Fourier pattern, $y_2$, exhibits a null at the center, in contrast to the plane wave diffracted pattern, $y_1$, which shows a central maximum.} In case of incoherent imaging with open and spiral phase apertures, we note that the generalized Wiener filters have nearly opposing polarity leading to a STED-like enhancement in the final reconstructed image \cite{singh2024high}. We further note that pure phase filters (spiral phase or potential other choices) are preferable in learning the diversity as intended here as they are energy conserving and simply redistribute the energy in the image (in case of incoherent systems) or the diffraction pattern (in case of coherent systems). Generalization of the proposed pseudo-data generation idea for multi-shot diversity with more than one pseudo-data may also be explored although this is not within the scope of our present work. The physically diverse pseudo-data generation as presented here can potentially improve the eventual performance of a range of imaging systems by helping them ease the ill-posedness of the underlying image reconstruction problem.

\section{Conclusion}
In this work, we introduced a novel framework for generating physically diverse pseudo-data to augment single-shot measurements, enabling robust solutions to the underlying inverse imaging problems {without requiring hardware modifications}. Both incoherent imaging and coherent imaging cases were presented and both the systems were simulated to use the spiral phase (placed in Fourier plane and illumination plane, respectively) as a diversity mechanism.
The raw image data pair ($y_1$, $y_2$) in both incoherent and coherent imaging examples is difficult to relate analytically. The data pair however has an implicit spatial correlation since the pair is generated with the same test object. The image-to-image translation network like UNet is observed to learn this implicit spatial relation by means of test examples and given a true data $y_1$, it is able to generate the pseudo-data $y'_2$ (corresponding to $y_2$) with a good accuracy. It is well-known from prior work in inverse problems that availability of $y'_2$ in addition to $y_1$ considerably reduces the ill-posedness of the inverse problem and the image reconstruction task may be performed with simpler algorithms with improved robustness. This demonstration thus opens up number of possibilities for the computational imaging community where the imaging system hardware can be further simplified while providing high quality images for downstream applications.

\section{Acknowledgements}
KK acknowledges partial support from Abdul Kalam Technology Innovation Fellowship (INAE) and the National Quantum Mission, DST, India. MK acknowledges support from Prime Minister Research Fellowship. JB acknowledges support from National Post Doctoral Fellowship from ANRF, India.
\section{References}
\bibliography{references}
\bibliographystyle{vancouver}
\end{document}